# Ferromagnetism above 1000 K in highly cation-ordered double-perovskite insulator $Sr_3OsO_6$

Yuki K. Wakabayashi[1], Yoshiharu Krockenberger[1], Naoto Tsujimoto[2], Tommy Boykin[1,†], Shinji Tsuneyuki[2], Yoshitaka Taniyasu[1] & Hideki Yamamoto[1]

[1] *NTT Basic Research Laboratories, NTT Corporation, Atsugi, Kanagawa 243-0198, Japan*
[2] *Department of Physics, University of Tokyo, 7-3-1 Hongo, Bunkyo-ku, Tokyo 113-0033, Japan*
[†] *on leave from University of Central Florida*

**Magnetic insulators have been intensively studied for over 100 years, and they, in particular ferrites, are considered to be the cradle of magnetic exchange interactions in solids. Their wide range of applications include microwave devices[1] and permanent magnets[2]. They are also suitable for spintronic devices owing to their high resistivity[3], low magnetic damping[4,5] and spin-dependent tunneling probabilities[6]. The Curie temperature ($T_C$) is the crucial factor determining the temperature range in which any ferri/ferromagnetic system remains stable. However, the record $T_C$ has stood for over eight decades in insulators and oxides (943 K for spinel ferrite $LiFe_5O_8$[7]). Here we show that a highly B-site ordered double-perovskite, $Sr_2(SrOs)O_6$ ($Sr_3OsO_6$), surpasses this long standing $T_C$ record by more than 100 K. We revealed this B-site ordering by atomic-resolution scanning transmission electron microscopy. The density functional theory (DFT) calculations suggest that the large spin-orbit coupling (SOC) of $Os^{6+}$ $5d^2$ orbitals drives the system toward a $J_{eff} = 3/2$ ferromagnetic (FM) insulating state[8-10]. Moreover, the $Sr_3OsO_6$ is the first epitaxially grown osmate, which means it is highly compatible with device fabrication processes and thus promising for spintronic applications.**

The B-site ordered double-perovskite $A_2BB'O_6$ family includes lots of fascinating magnetic materials such as half-metals[11-13], multiferroic materials[14], antiferromagnetic (AFM) materials[15] and magnetic insulators[8,16-18]. The A site is usually occupied by an alkaline-earth or rare-earth element, and B and B' are transition metal elements. Explorations of magnetism have mainly focused on varying the combination of transition metal elements at B and B' sites, and it has been believed that having them occupied by two different transition metal elements is a prerequisite for a magnetic order at high temperatures[12]. Some $4d$ or $5d$ element-containing double-perovskites, e.g., $Sr_2FeMoO_6$[11] ($T_C$ = 415 K), $Sr_2CrReO_6$[13] ($T_C$ = 634 K) and $Sr_2CrOsO_6$[17] ($T_C$ = 725 K), reach a point of FM instability at high temperatures, although the majority of double-perovskites show an AFM order or weak spin-glass behavior[19]. In contrast to those $4d$ or $5d$ double-perovskites that follow the above-mentioned criteria, we discovered FM ordering above 1000 K in a novel insulating double-perovskite $Sr_3OsO_6$, in which only one $5d$ transition metal element occupies the B sites. Remarkably, the $T_C$ of $Sr_3OsO_6$ (~1060 K) is about ten times higher than the previous highest magnetic transition temperature in double perovskites including only one transition element ($Sr_2ScOsO_6$, AFM, $T_N$ = 110 K[20]). The density functional theory (DFT) calculations suggest that the large SOC of $Os^{6+}$ $5d^2$ orbitals drives $Sr_3OsO_6$ toward the $J_{eff} = 3/2$ FM insulating state[8-10]. They also elucidate that the canted FM order has the lowest total energy.

High-quality B-site ordered double-perovskite $Sr_3OsO_6$ films (300-nm thick) were epitaxially grown on (001) $SrTiO_3$ substrates in a custom-designed molecular beam epitaxy (MBE) setup that provides a precise flux even for the high-melting-point element Os (3033°C) (METHODS). Maintaining a precise flux rate for each constituent cation (Os and Sr) with a simultaneous supply of $O_3$ is essential for avoiding deterioration of the magnetic properties; a deviation of only 2% from the optimal Os/Sr ratio is fatal (METHODS).

High-resolution scanning transmission electron microscopy (STEM) and transmission electron diffraction (TED), combined with high-resolution reciprocal space mapping (HRRSM) and reflection high-energy electron diffraction (RHEED), ascertained a cubic double-perovskite structure[12,16]. As schematically shown in Figs. 1a and 1d (viewed along [100] and [110] directions), Sr- or Os-occupied, fully Sr-occupied, fully Os-occupied and fully oxygen-occupied columns exist. The STEM images overtly demonstrate that these columns are arranged in a spatially ordered sequence. Since the intensity in the high-angle annular dark-field (HAADF)-STEM image is proportional to $\sim Z^n$ (n $\sim$ 1.7-2.0, and $Z$ is the atomic number), in Fig. 1b, the brightest and gray spheres are assigned to Sr- ($Z = 38$) or Os- ($Z = 76$) occupied and fully Sr-occupied columns, respectively. In Fig. 1e, the brightest and gray spheres are assigned to fully Os-occupied and fully Sr-occupied columns, respectively. In contrast to HAADF-STEM, annular bright-field (ABF)-STEM images allow for oxygen discrimination in addition to Sr and Os. Accordingly, fully oxygen-occupied columns were also detected, which are labeled O (insets of Figs. 1c and 1f). The energy dispersive X-ray spectroscopy (EDS)-STEM intensity profiles along the [100] direction shown in Figs. 1e and 1f complementarily confirm the above elemental assignments. The peak positions in the EDS profile of the Os $L$ shell (oxygen $K$ shell) agree well with the Os (O) positions determined by STEM. The STEM observation revealed the rock-salt type order of $Os^{6+}$ — the hexavalent state of Os is confirmed by X-ray photoemission spectroscopy (XPS) measurements (METHODS) — to an excellent extent, and this ordering is driven by the large difference in the electronic charges and ionic radii between $Sr^{2+}$ and $Os^{6+}$.[21] Accordingly, there are no Os-O-Os paths. Therefore, advanced mechanisms need to be considered as the Goodenough-Kanamori rules[22], which well predict magnetic interactions between two next-nearest-neighbor magnetic cations through a nonmagnetic anion, do not cover the theoretical framework exigent for the FM order in $Sr_3OsO_6$.

Figures 2a and 2b show the temperature dependence of magnetization versus the magnetic field (M-H) of a $Sr_3OsO_6$ film. The hysteretic response of the $Sr_3OsO_6$ film shows a soft magnetic behavior with the small coercive field of ~100 Oe at 1.9 K (Fig. 2b), and the saturation magnetization at 70000 Oe (Fig. 2a) decreases with increasing temperature. The saturation magnetization persists up to 1000 K [limit of measurement range (METHODS)], indicating $T_C >$ 1000 K. Figure 2c shows the magnetization versus temperature (M-T) curve with H = 2000 Oe. In Fig. 2b, we also plot the spontaneous magnetization as a function of temperature. The $T_C$ value, estimated from the extrapolation of the M-T curve to the zero magnetization axis, is about 1060 K (Fig. 2c). This is the highest $T_C$ among all insulators and oxides.

While such high $T_C$ is common for systems with free charge carriers, e.g., $Fe_3O_4$ and Co, their absence in $Sr_3OsO_6$ requires other exchange paths. The temperature dependence of resistivity ($\rho$) for a $Sr_3OsO_6$ film is shown in Fig. 3a. The electronic charge carriers [5d electrons in the $Os^{6+}$ state] move by hopping between localized electronic states, and this is supported by $\ln(\rho) \propto T^{-1/4}$ [variable range hopping (VRH) model] (Fig. 3b) along with the high resistivity value [$\rho$(300 K) = 75 $\Omega$cm]: other mechanisms, e.g., $\ln(\rho) \propto T^{-1/2}$ [Efros-Shklovskii Hopping (ESH) model] and $\ln(\rho) \propto T^{-1}$ [thermal activation (TA) model], are not supported by the electronic transport response. Figure 3c shows electron energy loss spectroscopy (EELS) spectra of a $Sr_3OsO_6$ film; we measured three different spots as indicated in the cross-sectional STEM image (inset of Fig. 3c) with a spot size of ~4 nm. An EELS spectrum corresponds to the loss function $Im(-1/\epsilon)$, where $\epsilon$ is a complex dielectric function. The three EELS spectra are almost identical, indicating that electronic states are uniform in the entire $Sr_3OsO_6$ film. The band gap (indicated by the black arrow), at which EELS intensities start to increase[23], is ~2.65 eV. These results indicate that $Sr_3OsO_6$ is an insulator with a band gap of ~2.65 eV. Accordingly, models based on the double exchange interaction and direct exchange interaction, where itinerant electrons are driving the magnetic order, can be ruled out as the origin of the emergent ferromagnetism.

The saturation magnetization of $Sr_3OsO_6$ (~49 emu/cc at 1.9 K) is significantly smaller than that for typical magnetic metals; e.g., $Nd_2Fe_{14}B$ (~1280 emu/cc), $SmCo_5$ (~860 emu/cc) and AlNiCo 5 (~1120 emu/cc)[24], and typical ferrites; e.g., $CoFe_2O_4$ (~430 emu/cc), $Y_3Fe_5O_{12}$ (~170 emu/cc) and $LiFe_5O_8$ (~390 emu/cc)[7]. The small saturation magnetization unique to $Sr_3OsO_6$ allows for small stray fields and low-energy spin-transfer-torque switching[25], which are advantageous for high-density-integration and low-power consumption spintronic devices. The small saturation magnetization is most likely associated with the low composition ratio of Os in $Sr_3OsO_6$. The saturation magnetic moment of Os at 1.9 K was estimated to be 0.77 $\mu_B$/Os, which is smaller than the expected value of the spin-only magnetic moment ($g\sqrt{S(S+1)}$ =2.83 $\mu_B$/Os) for the Os$^{6+}$ ($5d^2$ $t_{2g}^2$) state with $S$ = 1. This indicates that we have to take the SOC into account, which is often the case with 5$d$ systems[9,26], to understand the electronic states and physical properties of Sr$_3$OsO$_6$.

We analyzed the electronic and magnetic states of Sr$_3$OsO$_6$ by DFT with SOC (METHODS). It was revealed that the canted FM order (Fig. 4a) has the lowest total energy among all possible magnetic arrangements. However, the energy differences between the canted FM order and the collinear FM order (Extended Data Fig. 9a) or the AFM order (Extended Data Fig. 9b) are very small (~3.6 meV per atom or ~1.4 meV per atom, respectively), implying a competition among these orders. Note that such canted magnetic order is recently reported in other Os containing double perovskites[27,28]. The extended superexchange paths (Os-O-O-Os and Os-O-Sr-O-Os), which are well recognized to drive the magnetic order in Os containing double-perovskites[29,30], are one possible origin of the observed ferromagnetism in Sr$_3$OsO$_6$. Our GGA + $U$ + SOC calculation indicates that the electronic structure of Sr$_3$OsO$_6$ with the canted FM order has a gap (~0.37 eV) at the Fermi energy ($E_F$), equivalently an insulating state (Figs. 4b and 4c). With $U$ + SOC, the $t_{2g\uparrow}$ states are split into effective total angular momenta $J_{eff}$ = 3/2 (doublet) and $J_{eff}$ = 1/2 (singlet) states with opening a gap. The $J_{eff}$ = 3/2 states are fully occupied with two 5$d$ electrons per Os$^{6+}$, resulting in the insulating state. We note that a metallic grand state is obtained for the canted FM order when only SOC (without $U$) is taken into account. This is because the band dispersions of $J_{eff}$ = 3/2 and $J_{eff}$ = 1/2 states are greater than the spin-orbit splitting. The DFT calculation on the element-specific partial density-of-state (PDOS) (Fig. 4c) also revealed that the $J_{eff}$ = 3/2 and $J_{eff}$ = 1/2 bands around $E_F$ are mostly composed of the Os 5$d$ orbitals and the O 2$p$ component is very small, implying that $d$-$p$ hybridization is negligibly small. The calculated magnetic moment of osmium is 1.56 $\mu_B$/Os, which coincides better with the experimentally obtained value (0.77 $\mu_B$/Os) as compared with the spin-only magnetic moment (2.83 $\mu_B$/Os). Although further work is required to reveal the underlying mechanisms driving Sr$_3$OsO$_6$ into the robust FM order, and also to achieve a quantitative agreement of the band gap between experiment (~2.65 eV) and calculation (~0.37 eV), our calculation provides information on the magnetic arrangement at the grand state and how the energy gap is open by the interplay between electron correlation ($U$) and SOC.

Our current findings in epitaxial Sr$_3$OsO$_6$ films — an extraordinarily high $T_C$ of 1060 K, $J_{eff}$ = 3/2 insulating state, rock-salt type Os$^{6+}$ order and small magnetic moment — enrich the family of ferri/ferromagnetic insulators. Although the underlying electronic exchange mechanisms driving the robust FM order in Sr$_3$OsO$_6$ remain murky, applications of Sr$_3$OsO$_6$ to oxide-electronics beyond the current ferrite technology are feasible.

**Author Contributions** Y.K. and Y.K.W. prepared the samples. Y.K.W. performed experimental measurements and data analysis. N.T. and S.T. carried out the electronic-structure calculations. Y.K.W. wrote the paper. All authors contributed to the manuscript and the interpretation of the data.

**Acknowledgements** We thank Ken-ichi Sasaki for a valuable discussion. We thank Ai Ikeda for her help with the X-ray diffraction and resistivity measurements. We thank Hiroshi Irie for his help with the resistivity measurements. We thank Kazuhide Kumakura for his installation of MPMS3 SQUID-VSM oven option.

**Author information** Correspondence and requests for materials should be addressed to Y.K.W. (wakabayashi.yuki@lab.ntt.co.jp).

**References**

1. Naito, Y. & Suetake K. Application of ferrite to Electromagnetic wave absorber and its characteristics. IEEE Trans. Microw. Theory Tech. **19**, 65 (1973).
2. Tenaud, P., Morel, A., Kools, F., Le Breton, J. M. & Lechevallier, L. Recent improvement of hard ferrite permanent magnets based on La-Co substitution. J. Alloy. Comp. **370**, 331 (2004).
3. Avci, C. O. *et al*. Current-induced switching in a magnetic insulator. Nat. Mater. **16**, 309 (2017).
4. Kagiwara, Y. *et al*. Transmission of electrical signals by spin-wave interconversion in a magnetic insulator. Nature **464**, 262 (2010).
5. Schneider, T., Serga, A. A., Leven, B., Hillebrands, B., Stamps, R. L. & Kostylev, M. P. Realization of spin-wave logic gates. Appl. Phys. Lett. **92**, 022505 (2008).
6. Moodera, J. S., Hao, X., Gibson, G. A. & Meservey, R. Electron-spin polarization in tunnel junctions in zero applied field with ferromagnetic EuS barriers. Phys. Rev. Lett. **61**, 637 (1988).
7. Posnjak, E. & Barth, T. F. W. A New Type of Crystal Fine-Structure: Lithium Ferrite ($Li_2O·Fe_2O_3$). Phys. Rev. **38**, 2234 (1931).
8. Erickson, A. S. *et al*. Ferromagnetism in the Mott insulator $Ba_2NaOsO_6$. Phys. Rev. Lett. **99**, 016404 (2007).
9. Xiang, H. J. & Whangbo, M.-H. Cooperative Effect of Electron Correlation and Spin-Orbit Coupling on the Electronic and Magnetic Properties of $Ba_2NaOsO_6$. Phys. Rev. B **75**, 052407 (2007).
10. Gangopadhyay, S. & Pickett, W. E. Spin-orbit coupling, strong correlation, and insulator-metal transitions: The $J_{eff}$ = 3/2 ferromagnetic Dirac-Mott insulator $Ba_2NaOsO_6$. Phys. Rev. B **91**, 045133 (2015).
11. Kobayashi, K.-I., Kimura, T., Sawada, H., Terakura, K. & Tokura, Y. Room-temperature magnetoresistance in an oxide material with an ordered double-perovskite structure. Nature **395**, 677 (1998).
12. Chen, W.-t. *et al*. A half-metallic A- and B-site-ordered quadruple perovskite oxide $CaCu_3Fe_2Re_2O_{12}$ with large magnetization and a high transition temperature. Nat. commun. **5**, 3909 (2014).
13. Kato, H. *et al*. Metallic ordered double-perovskite $Sr_2CrReO_6$ with maximum Curie temperature of 635 K. Appl. Phys. Lett. **81**, 328 (2002).
14. Sakai, M. *et al*. Multiferroic thin film of $Bi_2NiMnO_6$ with ordered double-perovskite structure. Appl. Phys. Lett. **90**, 072903 (2007).
15. Yan, B. *et al*. Lattice-site-specific spin dynamics in double perovskite $Sr_2CoOsO_6$. Phys. Rev. Lett. **112**, 147202 (2014).
16. Iwasawa, H. *et al*. Strong correlation effects of the Re $5d$ electrons on the metal-insulator transition in $Ca_2FeReO_6$. Phys. Rev. B **71**, 075106 (2005).
17. Krockenberger, Y. *et al*. $Sr_2CrOsO_6$: End point of a spin-polarized metal-insulator transition by $5d$ band filling. Phys. Rev. B **75**, 020404(R) (2007).
18. Feng, H. L. *et al*. $Ba_2NiOsO_6$: A Dirac-Mott insulator with ferromagnetism near 100 K. Phys. Rev. B **94**, 235158 (2016).
19. Battle, P. D., Gibb, T. C., Herod, A. J., Kim, S.-H. & Munns, P. H. Investigation of magnetic frustration in $A_2FeMO_6$ (A = Ca, Sr, Ba; M = Nb, Ta, Sb) by magnetometry and mossbauer Spectroscopy. J. Mater. Chem. **5**, 865 (1995).
20. Taylor, A. E. et al. Magnetic order and electronic structure of the $5d^3$ double perovskite $Sr_2ScOsO_6$. Phys. Rev. B **91**, 100406 (2015).
21. Feng, H. L. *et al*. High-pressure crystal growth and electromagnetic properties of 5d double-perovskite $Ca_3OsO_6$. J. Solid State Chem. **201**, 186 (2013).
22. Goodenough, J. B. Magnetism and the chemical bond (Wiley, New York, 1963).

23  Stoughton, S. *et al*. Adsorption-controlled growth of BiVO$_4$ by molecular-beam epitaxy. Appl. Phys. Lett. Mater. **1**, 42112 (2013).
24  Coey, J. M. D. Hard Magnetic Materials: A perspective. IEEE Trans. Magn. **47**, 4671 (2011)
25  Stiles, M. D. & Zangwill, A. Anatomy of spin-transfer torque. Phys. Rev. B **66**, 014407 (2002).
26  Kim, B. J. *et al*. Phase-sensitive observation of a spin-orbital Mott state in Sr$_2$IrO$_4$. Science. **323**, 1329 (2009).
27  Feng, H. L. *et al*. Canted ferrimagnetism and giant coercivity in the nonstoichiometric double perovskite La$_2$Ni$_{1.19}$Os$_{0.81}$O$_6$. Phys. Rev. B **97,** 184407 (2018).
28  Yan, B. *et al*. Lattice-Site-Specific Spin Dynamics in Double Perovskite Sr$_2$CoOsO$_6$. Phys. Rev. Lett. 112, 147202 (2014).
29  Shi, Y. *et al*. Crystal Growth and Structure and Magnetic Properties of the 5d Oxide Ca$_3$LiOsO$_6$: Extended Superexchange Magnetic Interaction in Oxide. J. Am. Chem. Soc. **132**, 8474–8483 (2010).
30  Kanungo, S., Yan, B., Felser, C. & Jansen, M. Active role of nonmagnetic cations in magnetic interactions for double-perovskite Sr$_2$*B*OsO$_6$ (*B* = Y,In,Sc). Phys. Rev. B **93**, 161116 (2016).

**Figures and figure legends**

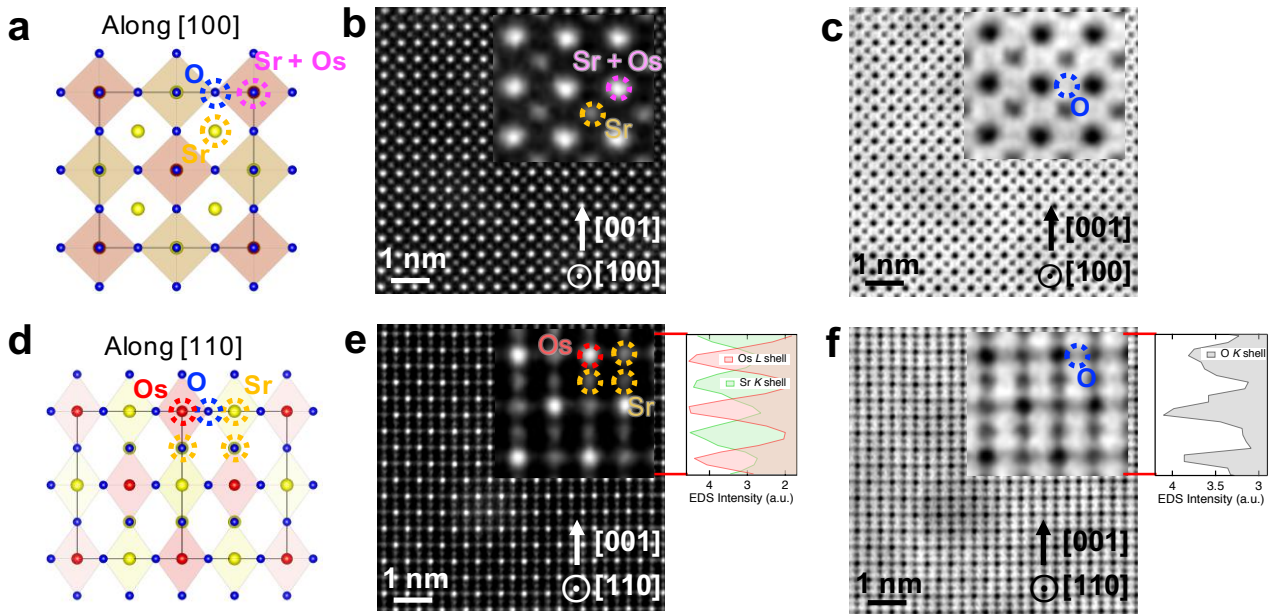

**Figure 1 Atomic-resolution STEM images of a $Sr_3OsO_6$ film. a**, Schematic diagram of the $Sr_3OsO_6$ viewed along the [100] direction. **b**, **c**, HAADF-STEM (**b**) and ABF-STEM (**c**) images near the center of the $Sr_3OsO_6$ layer along the [100] direction. **d**, Schematic diagram of the $Sr_3OsO_6$ viewed along the [110] direction. **e**, **f**, HAADF-STEM (**e**) and ABF-STEM (**f**) images near the center of the $Sr_3OsO_6$ layer along the [110] direction. The insets in **e** and **f** show enlarged views, and the corresponding graphs show EDS-STEM intensity profiles along the [001] direction. In all figures, purple, yellow, red and blue dotted circles indicate Sr- or Os- occupied, fully Sr-occupied, fully Os-occupied and fully oxygen-occupied columns, respectively.

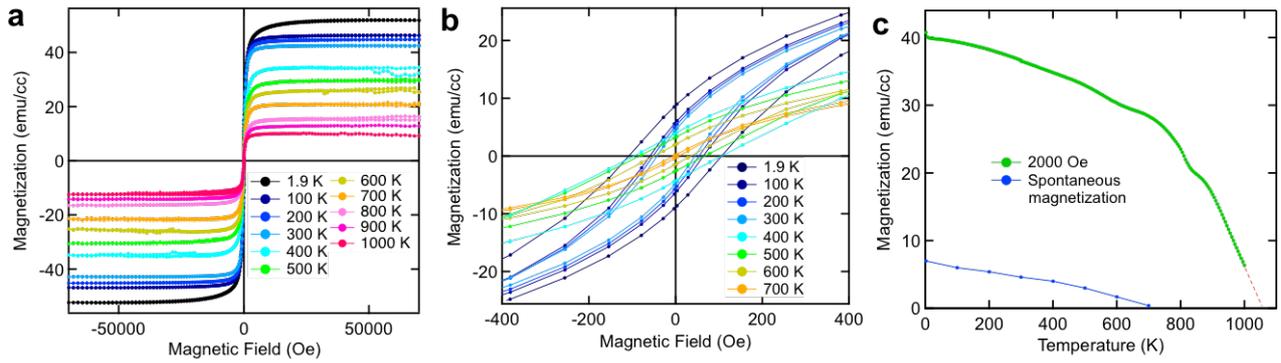

**Figure 2 Magnetic properties of a $Sr_3OsO_6$ film. a**, In-plane *M-H* curves at 1.9 to 1000 K for a $Sr_3OsO_6$ film. Here, *H* was applied to the [100] direction. **b**, Close-up near the zero magnetic field in **a**. **c**, *M-T* curve with *H* = 2000 Oe applied to the [100] direction for a $Sr_3OsO_6$ film. Spontaneous magnetization deduced from Fig. 3(b) as a function of temperature is also shown.

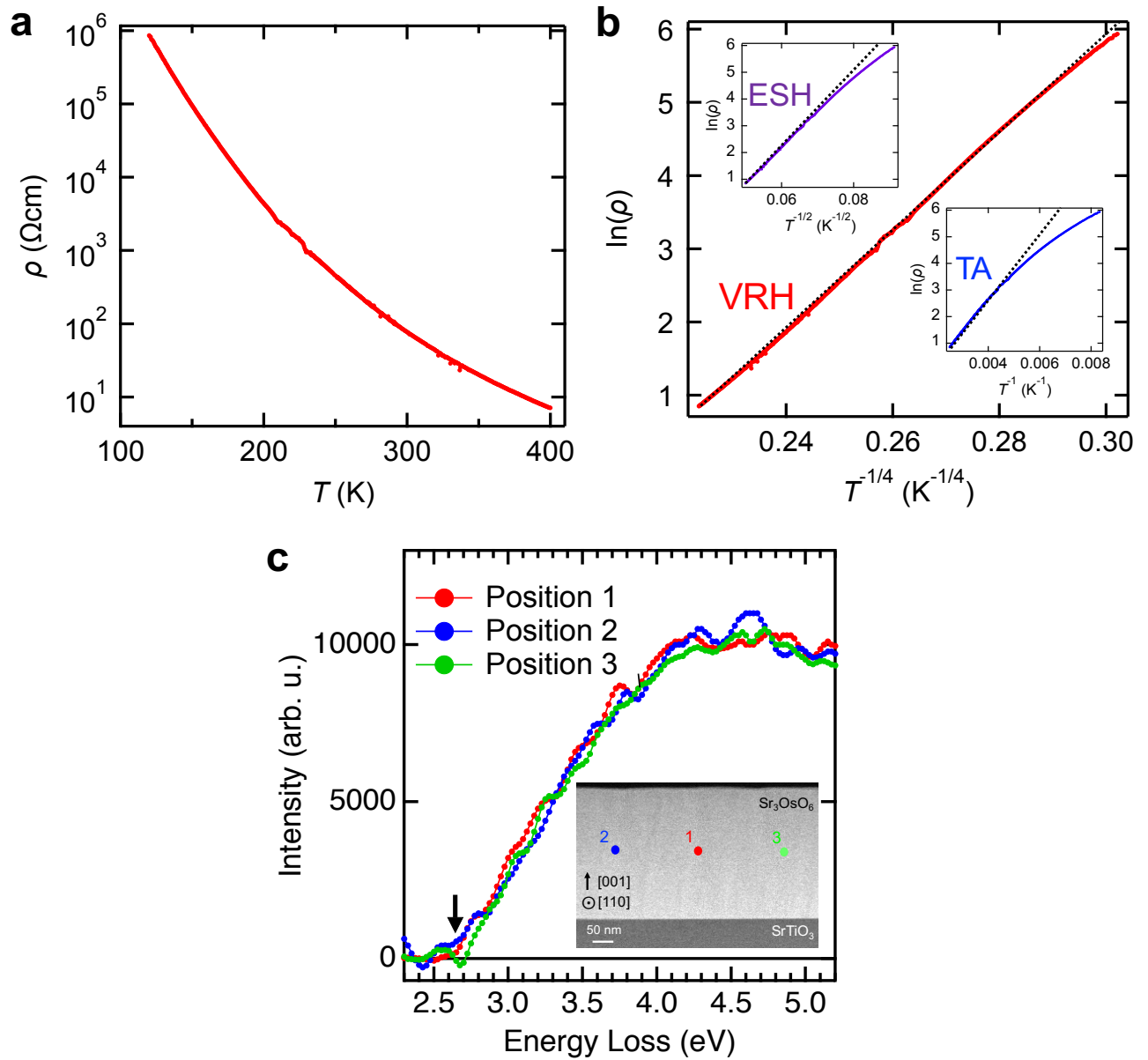

**Figure 3. Resistivity and dielectric properties of a $Sr_3OsO_6$ film. a**, $\rho$-$T$ curve for a $Sr_3OsO_6$ film. **b**, Logarithm of $\rho$ versus $T^{1/4}$ plot, corresponding to the VRH model. The insets of **b** show the logarithm of $\rho$ versus $T^{1/2}$ and $T^{-1}$ plots, corresponding to the ESH and TA models, respectively. The black dashed lines in **b** are guides for the eye. **c**, The EELS spectra of a $Sr_3OsO_6$ film measured at the spots indicated in the cross-sectional STEM image (inset). The background was corrected with a power-law fit from 2 to 2.3 eV.

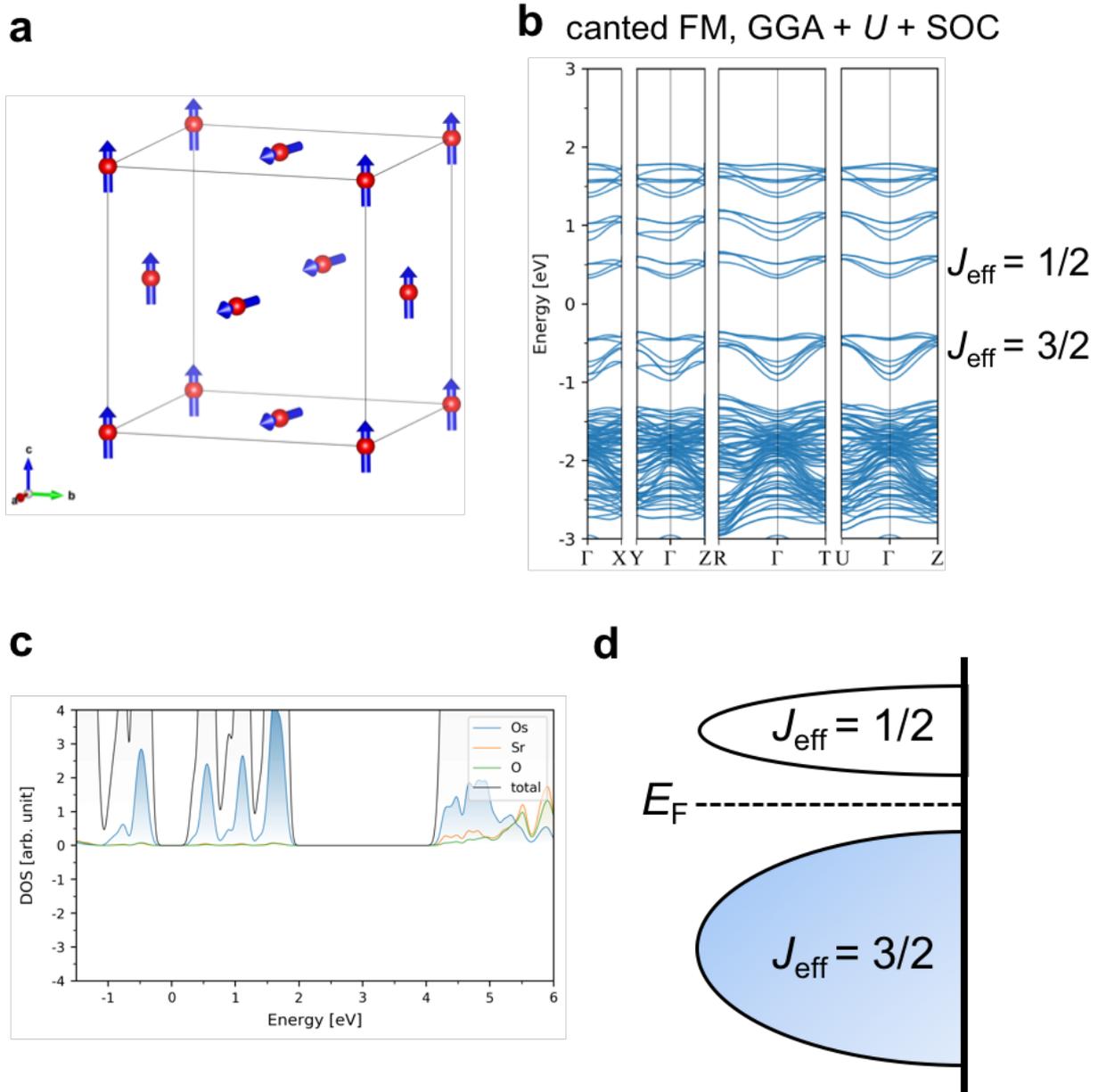

**Figure 4. Electronic-structure calculations. a**, Schematic diagram of the magnetic ground state (canted FM order) of $Sr_3OsO_6$ obtained from the DFT calculation. In **a**, red spheres and blue arrows indicate Os atoms and magnetic moments of Os atoms, respectively, and the Sr and O atoms are omitted for the simplicity. **b**, The band structures for $Sr_3OsO_6$ with the canted FM order calculated by GGA + $U$ + SOC. **c**, The element-specific partial density-of-state (PDOS) for the canted FM order calculated by GGA + $U$ + SOC. **d**, Schematic energy diagrams for the Os $5d^2$ configurations. In **d**, only PDOS for Os is taken into account and the contributions by Sr and O are omitted for the simplicity.

**METHODS**

**Growth of $Sr_3OsO_6$ on $SrTiO_3$.** We grew the high-quality epitaxial B-site ordered double-perovskite (001) $Sr_3OsO_6$ films (300- or 250-nm thick) on (001) $SrTiO_3$ substrates (CrysTec GmbH) in a custom designed molecular beam epitaxy (MBE) system[31,32] (Extended Data Fig. 1a). After cleaning with $CHCl_3$ (10 min, 2 times) and acetone (5 min) in ultrasonic cleaner, the $SrTiO_3$ substrate was introduced in the MBE growth chamber. After degassing the substrate at 400°C for 30 minutes and successive thermal cleaning at 650°C for 30 min, we grew a $Sr_3OsO_6$ film. The growth temperature was 650°C. The oxidation during the growth was carried out with $O_3$ gas (non-distilled, ~10% concentration) from a commercial ozone generator, and the resultant chamber pressure during the growth was ~4.5 × $10^{-6}$ Torr. After the growth, films were cooled to room temperature under ultra-high vacuum (UHV). The MBE system is equipped with multiple e-beam evaporators (Hydra, Thermionics) for Sr and Os. The electron impact emission spectroscopy (EIES) sensor (Guardian, Inficon) is located next to the sample holder in the same horizontal plane. The sensor head is equipped with a filament, which generates thermal electrons for the excitation of Sr and Os atoms. Optical band-pass filters are used for element-specific detection of the excited optical signals, since the emitted light spectra are characteristic for Sr and Os. The EIES sensor is equipped with photomultipliers (PMTs) located outside of the vacuum chamber that amplify optical signals. The Sr and Os fluxes measured by EIES were kept constant (Extended Data Fig. 1b) by the proportional-integral-derivative (PID) control of the evaporation source power supply. We optimized the flux ratio of Sr and Os to obtain a $Sr_3OsO_6$ film with a high saturation magnetic moment. Extended Data Fig. 2 shows the in-plane *M-H* curves at 300 K for $Sr_3OsO_6$ films grown with a different flux ratio of Sr and Os. The saturation magnetic moment of the film grown with the flux ratio of Sr:Os = 2.05:1 is ten or more times larger than those for the films grown with the flux ratio of Sr:Os = 2.05:1.02 and 2.05:0.98. This means that the magnetic properties of $Sr_3OsO_6$ films are very sensitive to the Sr/Os ratio and that well-controlled Sr and Os fluxes during the growth are important for the high saturation magnetic moment. Therefore, in this study, we fixed the flux ratio of Sr:Os = 2.05:1.

The cubic crystal structure of $Sr_3OsO_6$ is illustrated in Extended Data Fig. 3a. Extended Data Figs. 3b and 3c show reflection high-energy electron diffraction (RHEED) patterns of a $Sr_3OsO_6$ thin film surface, where the sharp streaky patterns with clear surface reconstruction indicate the growth of the $Sr_3OsO_6$ film proceeded in a layer-by-layer manner, leading to the high crystalline quality of the film. Notably, [01*l*] diffractions are not seen (Extended Data Fig. 3b) due to extinction rules, indicating the formation of a cubic B-site ordered double-perovskite[12,16]. The cubic structure model is further evidenced by high-resolution X-ray reciprocal space mapping (HRRSM) (Extended Data Fig. 3d): the in-plane and out-of-plane lattice constants of $Sr_3OsO_6$ are identical within the resolution limits (8.24±0.03 Å and 8.22±0.03 Å, respectively). It is therefore reasonable that the $Sr_3OsO_6$ films are epitaxially but not coherently grown on the $SrTiO_3$ (3.905 Å) substrate.

**Transmission electron microscopy.** High-angle annular dark-field (HAADF) and annular bright-field (ABF) scanning transmission electron microscopy (STEM) images were taken with a JEOL JEM-ARM 200F microscope. Electron energy loss spectroscopy (EELS) spectra of a $Sr_3OsO_6$ film were recorded from a spot with ~4 nm diameter also with a JEOL JEM-ARM 200F microscope.

Extended Data Figs. 3e-3g show cross-sectional high-angle annular dark-field scanning transmission electron microscopy (HAADF-STEM) images of a $Sr_3OsO_6$ film taken along the [100] direction. Since the intensity in the HAADF image is proportional to $\sim Z^n$ (n ~ 1.7-2.0, and *Z* is the atomic number)[33], the HAADF intensity is dominated by Sr (*Z* = 38) and Os (*Z* = 76) ions, whereas oxygen (*Z* = 8) is scarcely discernible. At a glance one can recognize that a single-crystalline $Sr_3OsO_6$ film with an abrupt substrate/film interface has been grown epitaxially on a (001) $SrTiO_3$ substrate, as expected from the RHEED. Misfit dislocations at the $Sr_3OsO_6/SrTiO_3$ interface (Extended Data Fig. 3g) are due to the ~5% larger lattice constant of the perovskite

Sr$_3$OsO$_6$ lattice (8.23 Å/2 = 4.115 Å) than that of SrTiO$_3$ (3.905 Å). The cubic crystal structure of Sr$_3$OsO$_6$ was also confirmed by the STEM analysis. In addition to the [100] direction (Extended Data Figs. 3e-3g), the epitaxial growth of the Sr$_3$OsO$_6$ layer on the SrTiO$_3$ substrate was also confirmed in STEM images taken along the [110] direction (Extended Data Figs. 4a-4c). The rock-salt type order of Os$^{6+}$ confirmed in the main text (Fig. 1) is observed to an excellent extent (Extended Data Figs. 4d and 4e).

**Chemical composition of a Sr$_3$OsO$_6$ film.** Extended Data Fig. 5a shows the depth profile of the chemical composition of a Sr$_3$OsO$_6$ film (250-nm thick) estimated from Rutherford backscattering spectroscopy (RBS). The chemical composition of the Sr$_3$OsO$_6$ layer is uniform (Sr:Os:O = 2.7±0.1:1.15±0.05:6.15±0.4). The concentrations of Os and oxygen are slightly larger than those for an ideal composition (Sr:Os:O = 3:1:6). This slight difference may originate from the non-stoichiometry and existence of the very small amount of paramagnetic OsO$_2$[34], which was observed in the X-ray diffraction (XRD) measurements, as described later. To exclude the possibility of the contamination by magnetic impurities, we performed EDS measurement for a Sr$_3$OsO$_6$ film (Extended Data Fig. 5b). There are no peaks except for Sr, Os, Ti, oxygen and C, which confirms the absence of magnetic impurities.

**X-ray diffraction.** $2\theta$-$\theta$ and reciprocal space map XRD measurements of the Sr$_3$OsO$_6$ films were performed with a Bruker D8 diffractometer using monochromatic Cu K$\alpha_1$ radiation at room temperature. In Extended Data Fig. 5c, we show the $\theta$-$2\theta$ XRD pattern for a Sr$_3$OsO$_6$ film. In addition to the diffraction peaks of the SrTiO$_3$ substrates, (002) and (004) diffractions from Sr$_3$OsO$_6$ are clearly observed. (001) and (003) diffractions from Sr$_3$OsO$_6$ are not seen due to the extinction rules. Note that traces of OsO$_2$, which is known as a paramagnetic metal[35], are detected as indicated by *. The XRD intensities of OsO$_2$ is about 700 times smaller than those of Sr$_3$OsO$_6$ and segregation of OsO$_2$ is not discernible in the STEM images, indicating that volume fraction of OsO$_2$ (paramagnetic metal) is negligibly small. Therefore, Sr$_3$OsO$_6$ predominates the magnetic response of the film.

**X-ray photoemission measurements.** XPS is one of the most powerful methods to determine the valence of Os in compounds[36,37], since the $4f_{7/2}$ core level binding energies in Os compounds with well-defined oxidation states are known. ULVAC-PHI Model XPS5700 with a monochromatized Al K$\alpha$ (1486.6 eV) source operated at 200 W was used for the experiment. The scale of binding energy was calibrated against the C 1$s$ line (284.6 eV). Extended Data Fig. 6 shows the Os $4f$ spectrum of a Sr$_3$OsO$_6$ film at 300 K. The observed $4f_{7/2}$ binding energy (54.1 eV) is close to the reported values for those of Os$^{6+}$ states (53.2-53.8 eV) and far from those of Os$^{2+}$ states (49.7 eV), Os$^{3+}$ states (50.4-51.0 eV), Os$^{4+}$ states (51.7-52.3 eV) and Os$^{8+}$ states (55.9-56.3 eV)[36,37]. Accordingly, the hexavalent state of Os (Os$^{6+}$) is supported. Note that a shoulder structure at ~ 53 eV may originate from a surface layer formed due to the slightly hygroscopic nature of Sr$_3$OsO$_6$ as the sample was transferred to the XPS apparatus not *in vacuo* but in atmosphere.

**Resistivity measurements.** Resistivity was measured using the four-probe method in a Physical Property Measurement System (PPMS) Dynacool sample chamber. The Ag electrodes deposited on a Sr$_3$OsO$_6$ surface were connected to an Agilent 3458A Multimeter.

**Magnetic measurements.** The magnetization measurements for Sr$_3$OsO$_6$ films were performed with a Quantum Design MPMS3 SQUID-VSM magnetometer. Using a quartz sample holder (oven sample holder), we measured the *M-T* curves with increasing temperature from 1.9 (300) to 300 (1000) K with $H$ = 2000 Oe applied along the [100] direction. In the *M-T* measurements, *M* was measured with increasing temperature after the sample was cooled to 1.9 (300) K from 300 (1000) K without a magnetic field. We also measured *M-H* curves at 1.9-300 K (400-1000 K) using the quartz sample holder (oven sample holder).

To check the accuracy of the measurement temperature in the MPMS SQUID-VSM magnetometer, we measured the magnetic properties of a pure Ni reference plate (Quantum Design Part Number: 4505-155). The *M-H* curves at 300 and 1000 K show FM and paramagnetic response, respectively (Extended Data Fig. 7a), and the magnetization of Ni rapidly increases between 623 and 629 K (Extended Data Fig. 7b). These results indicate that $T_C$ of Ni is between 623 and 629 K. This is consistent with the $T_C$ value in the literature (627 K)[38]. Thus, the error of the measurement temperature in the MPMS SQUID-VSM magnetometer is less than $\pm$ 4 K.

Extended Data Fig. 7c shows the in-plane $M$-$H$ curves at 1.9, 300 and 1000 K for a $SrTiO_3$ substrate. They show only a linear diamagnetic response at 300 and 1000 K. The nonlinear magnetic response near the zero magnetic field at 1.9 K indicates the existence of a tiny amount of a paramagnetic impurity in the $SrTiO_3$ substrate. In Figs. 2a and 2b, the linear diamagnetic response of the magnetic moment for the $SrTiO_3$ substrate was subtracted from the raw $M$ data.

Extended Data Fig. 7d shows the $M$-$T$ curve with $H$ = 2000 Oe for the oven sample holder without a sample. The curve shows a dip structure at around 800 K. This means that the dip structure in the $M$-$T$ curve at around 800 K for the $Sr_3OsO_6$ film (Fig. 2c) is an unavoidable experimental artifact.

The magnetic properties of $Sr_3OsO_6$ at 300 K did not change much after it was heated to 1000 K as shown in Fig. 8a. This means that heating to 1000 K does not affect much its magnetic properties.

Although the $Sr_3OsO_6$ films were epitaxially grown on the $SrTiO_3$ substrate, the shapes of the in-plane $M$-$H$ curves measured with $H$ applied to the [100] and [110] directions are identical (Extended Data Fig. 8b). This indicates that the in-plane magnetic anisotropy of the $Sr_3OsO_6$ film is negligibly small. This small magnetic anisotropy might be related to the misfit dislocations (Extended Data Figs. 3e-3g), which often decrease the magnetic anisotropy of magnetic insulators[39,40].

**The electronic-structure calculations.** The electronic-structure calculations were based on density functional theory (DFT). The calculations were performed by using Vienna *Ab initio* Simulation Package (VASP)[41,42] with the projector augmented-wave (PAW)[43,44] method and the Parder-Burke-Ernzerhof (PBE)[45] functional of the generalized gradient approximation (GGA)[46]. To describe the localization of Os 5$d$ electrons accurately, we used the DFT + $U$ calculations[47]. The value of the screened Coulomb interaction $U$ = 3 eV was used for the Os atoms. This value is comparable to reported values for Os containing double perovskites (2-4 eV)[9,10,18,48]. The contribution of the spin orbit coupling was included in our calculations. The crystal structure was optimized for the conventional unit cell (40 atoms) of $Sr_3OsO_6$ whose lattice constant was fixed to the experimental value 8.23 Å. We performed the optimization until all forces on the atoms become smaller than $10^{-5}$ eV/Å with a $\Gamma$-centered $2\times2\times2$ $k$-point grid and cut-off energy of 800 eV. The total energies and electronic structures are calculated with the optimized crystal structures resulting in the canted FM ground state (Fig. 4a).

By comparing with the total energies of the magnetic ground state (canted FM order) (Fig. 4a), collinear FM order [Extended Data Fig. 9(a)] and the AFM order [Extended Data Fig. 9(b)], we found that the energy differences between the canted FM order and the collinear FM order, and between the canted FM order and the AFM order are very small (~3.6 meV per atom and ~1.4 meV per atom, respectively), implying a competition among these orders.

**References for METHODS**
31 Naito, M. & Sato, H. Stoichiometry control of atomic beam fluxes by precipitated impurity phase detection in growth of $(Pr,Ce)_2CuO_4$ and $(La,Sr)_2CuO_4$ films. Appl. Phys. Lett. **67**, 2557 (1995).
32 Yamamoto, H., Krockenberger, Y. & Naito, M. Multi-source MBE with high-precision rate control system as a synthesis method sui generis for multi-cation metal oxides. J. Cryst. Growth **378**, 184 (2013).
33 Nellist, P. D. & Pennycook, S. J. Advances in Imaging and Electron Physics p. 147 (Elsevier, 2000).
34 Greedan, J. E., Willson, D. B. & Haas, T. E. The metallic nature of osmium dioxide. Inorg. Chem.

**7,** 2461 (1968).
35  Yen, P. C., Chen, R. S., Chen, C. C., Huang, Y. S. & Tiong, K. K. Growth and characterization of $OsO_2$ single crystals. J. Cryst. Growth **262,** 271 (2004).
36  White, D. L., Andrews, S. B., Faller, J. W. & Barrnett, R. J. The chemical nature of osmium tetroxide fixation and staining of membranes by X-ray photoelectron spectroscopy. Biochim. Biophys. Acta **436,** 577–592 (1976).
37  Hayakawa, Y. *et al.* X-ray photoelectron spectroscopy of highly conducting and amorphous osmium dioxide thin films. Thin Solid Films **347,** 56–59 (1999).
38  Kouvel, J. S. & Fisher, M. E. Detailed magnetic behavior of nickel near its Curie temperature. Phys. Rev. **136,** 1626 (1964).
39  Margulies, D. T. *et al*. Origin of the anomalous magnetic behavior in single crystal $Fe_3O_4$ films. Phys. Rev. Lett. **79,** 5162 (1997).
40  Wakabayashi, Y. K. *et al*. Electronic structure and magnetic properties of magnetically dead layers in epitaxial $CoFe_2O_4/Al_2O_3/Si(111)$ films studied by X-ray magnetic circular dichroism. Phys. Rev. B **96,** 104410 (2017).
41  Kresse, G. & Hafner, J. *Ab initio* molecular dynamics for liquid metals. Physical Review B **47,** 558 (1993).
42  Kresse, G. & Furthmüller, J. Efficient iterative schemes for *ab initio* total-energy calculations using a plane-wave basis set. Physical Review B **54,** 11169 (1996).
43  Blochl, P. E. Projector augmented-wave method. Phys. Rev. B **50,** 17953 (1994).
44  Kresse, G. & Joubert, D. From ultrasoft pseudopotentials to the projector augmented-wave method. Phys. Rev. B **59,** 1758 (1999).
45  Perdew, J. P., Burke, K. & Ernzerhof, M. Generalized Gradient Approximation Made Simple. Phys. Rev. Lett. **77,** 3865 (1996).
46  Perdew, J. P. Accurate Density Functional for the Energy: Real-Space Cutoff of the Gradient Expansion for the Exchange Hole. Phys. Rev. Lett. **55,** 1665 (1985).
47  Dudarev, S. L., Savrasov, S. Y., Humphreys, C. J. & Sutton, a. P. Electron-energy-loss spectra and the structural stability of nickel oxide: An LSDA+U study. Phys. Rev. B **57,** 1505 (1998).
48  Kanungo, S., Yan, B., Felser, C. & Jansen, M. Active role of nonmagnetic cations in magnetic interactions for double-perovskite $Sr_2BOsO_6$ (*B* = Y,In,Sc). Phys. Rev. B **93,** 161116 (2016).

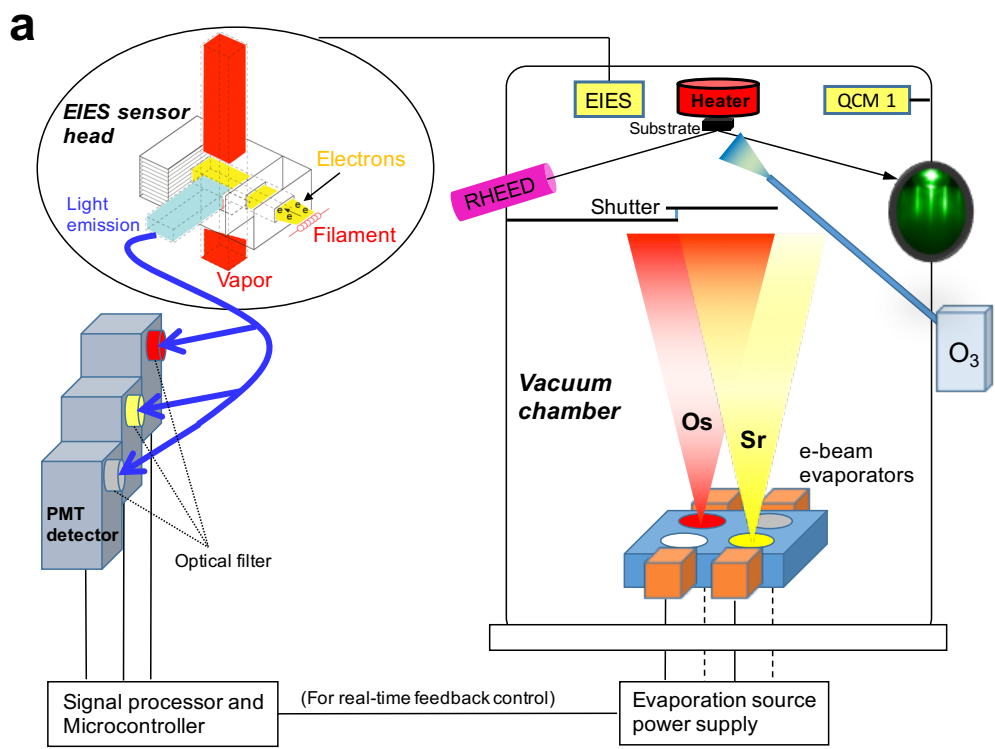

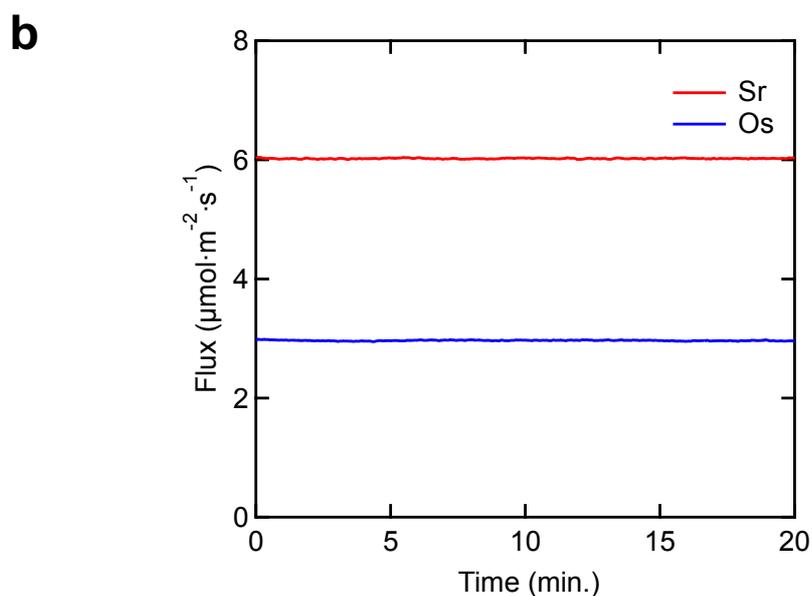

**Extended Data Figure 1 Multi-source oxide MBE setup and fluxes. a**, Schematic illustration of our multi-source oxide MBE setup. EIES: Electron Impact Emission Spectroscopy. QCM: Quartz Crystal Microbalance. RHEED: Reflection High-Energy Electron Diffraction. PMT: Photomultiplier Tube. **b**, Sr and Os fluxes measured by EIES during the growth.

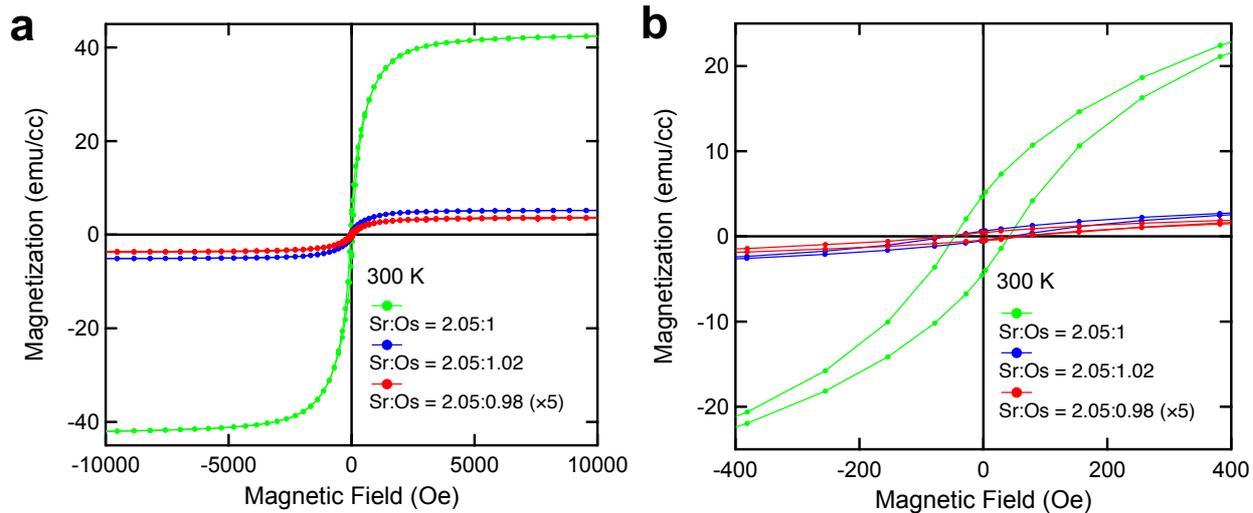

**Extended Data Figure 2 Magnetic properties of $Sr_3OsO_6$ films grown with a different flux ratio of Sr and Os. a**, In-plane *M-H* curves at 300 K for $Sr_3OsO_6$ films grown with a different flux ratio of Sr and Os. Here, *H* was applied to the [100] direction. **b**, Close-up near the zero magnetic field in **a**.

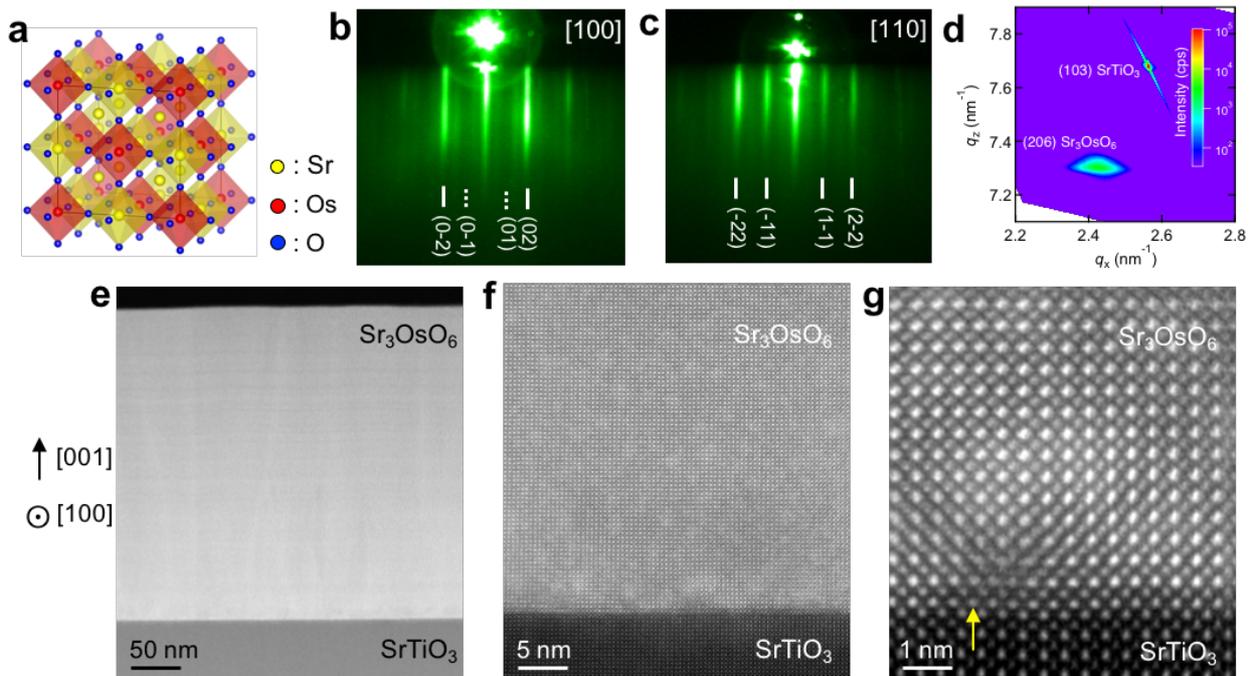

**Extended Data Figure 3 Crystal structure, RHEED, X-ray HRRSM and HAADF-STEM for a Sr$_3$OsO$_6$ film. a**, Schematic diagram of the B-site ordered double-perovskite Sr$_3$OsO$_6$. The yellow, red and blue spheres indicate Sr, Os and O ions, respectively. **b, c,** RHEED patterns of an epitaxial Sr$_3$OsO$_6$ film surface, where the incident electron beams are parallel to [100] (**b**) and [110] (**c**). **d**, X-ray HRRSM of a Sr$_3$OsO$_6$ film around the SrTiO$_3$ (103) reflection. **e,** Cross-sectional HAADF-STEM images of a Sr$_3$OsO$_6$ film taken along the [100] direction. **f,** Magnified image near the interface in **e**. **g,** Magnified image near the interface in **f**. In **g**, a misfit dislocation at the Sr$_3$OsO$_6$/SrTiO$_3$ interface is indicated by the yellow arrow.

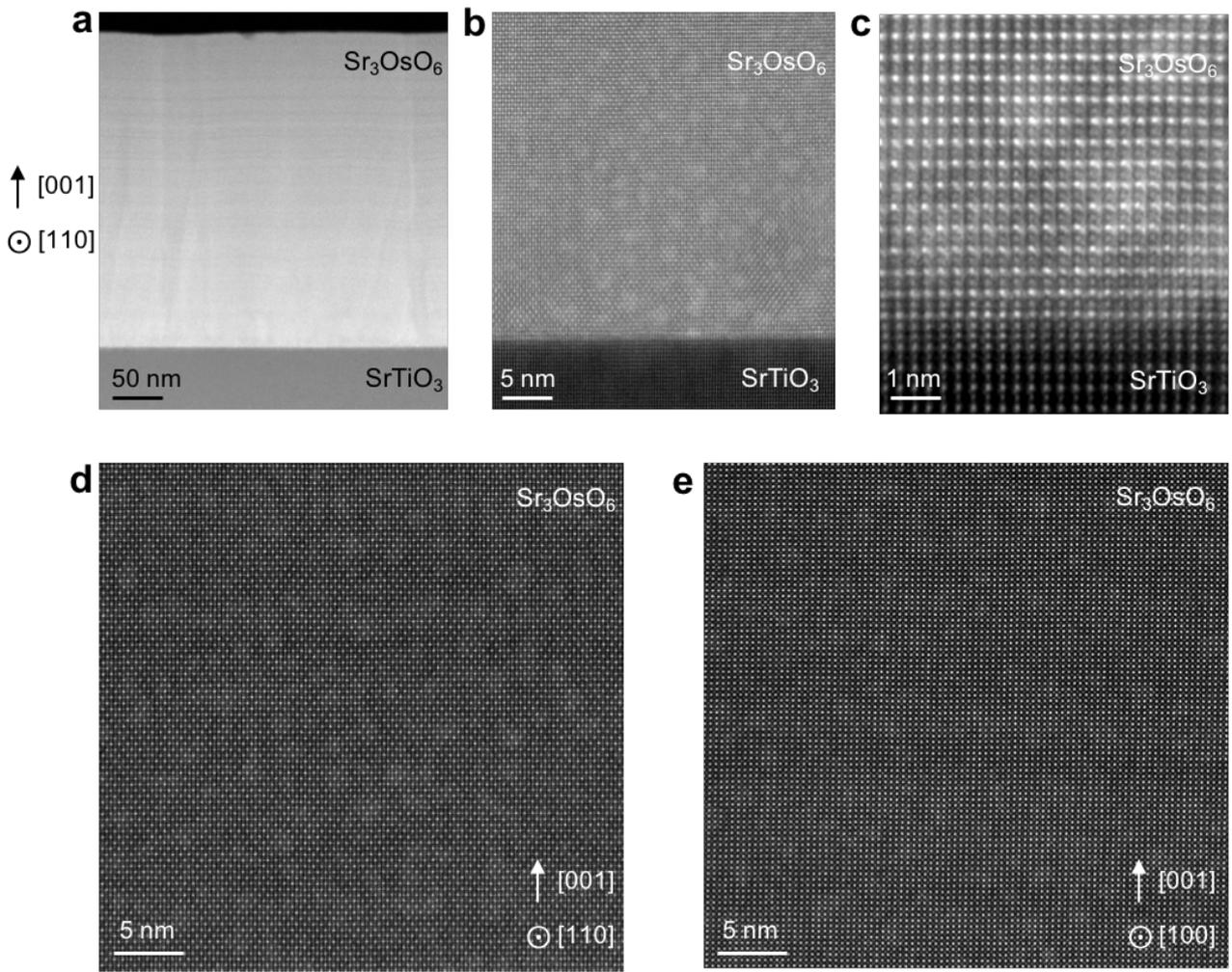

**Extended Data Figure 4 STEM images for a Sr$_3$OsO$_6$ film. a,** Cross-sectional HAADF-STEM image of a Sr$_3$OsO$_6$ film taken along the [110] direction. **b,** Magnified image near the interface in **a**. **c,** Magnified image near the interface in **b**. **d, e**, HAADF-STEM images of the Sr$_3$OsO$_6$ layer taken along the [110] (**d**) and [100] (**e**) directions.

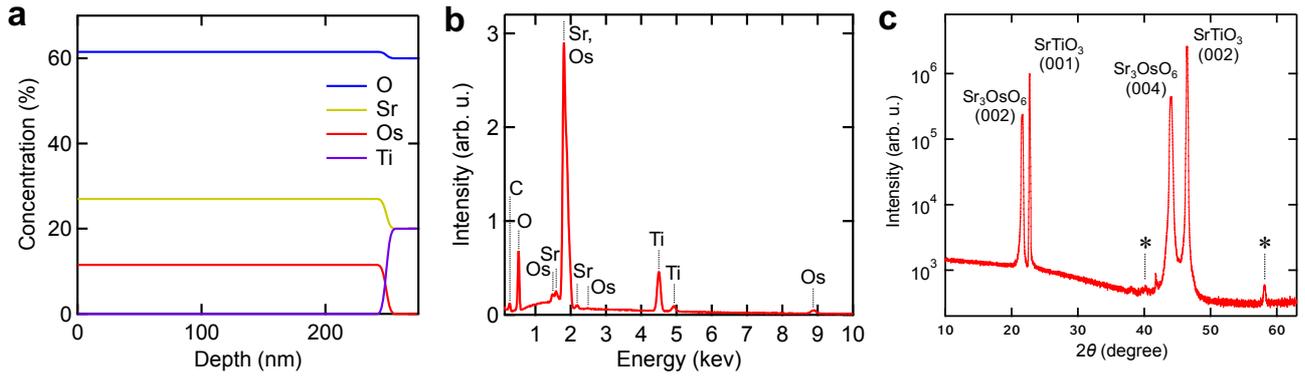

**Extended Data Figure 5 Chemical composition, EDS spectrum and $\theta$-$2\theta$ XRD pattern of $Sr_3OsO_6$ films. a**, Depth profile of the chemical composition of a $Sr_3OsO_6$ film (250-nm thick) estimated from RBS. **b**, EDS spectrum of a $Sr_3OsO_6$ film, which was taken from a wide area (1 × 1 mm$^2$). **c**, $\theta$-$2\theta$ XRD pattern for a $Sr_3OsO_6$ film. Traces of $OsO_2$ are detected as indicated by ✶.

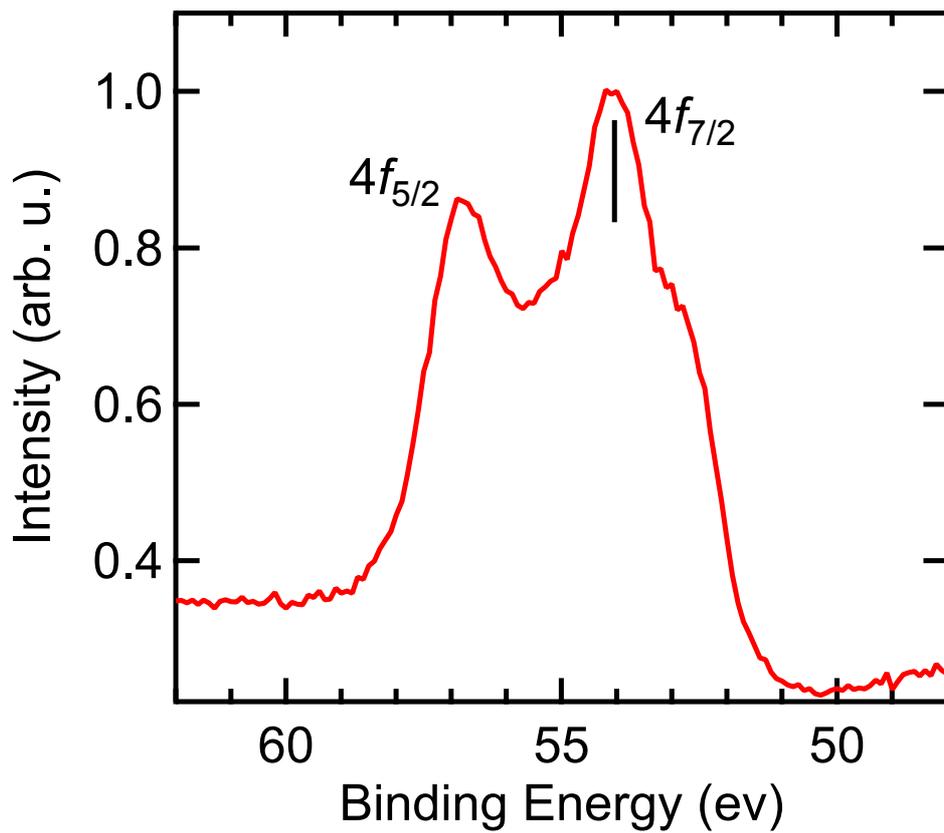

**Extended Data Figure 6 XPS of a Sr$_3$OsO$_6$ film.** The Os 4$f$ XPS spectrum of a Sr$_3$OsO$_6$ film at 300 K.

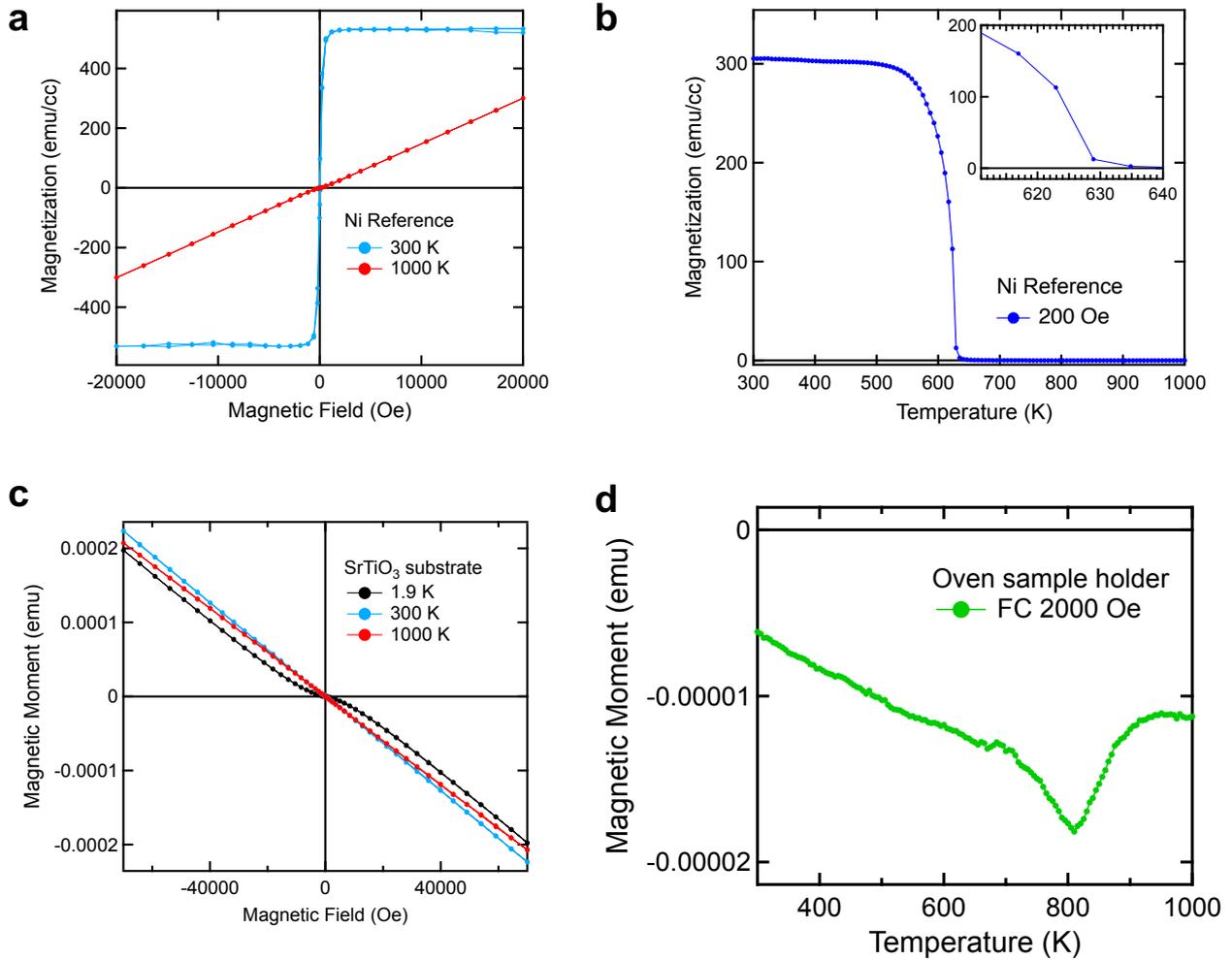

**Extended Data Figure 7 Magnetic properties of a Ni reference plate and a SrTiO$_3$ substrate, and experimental artifact from oven sample holder. a**, *M-H* curve at 300 and 1000 K for a Ni plate. Here, *H* was applied to the in-plane direction. **b**, *M-T* curve with *H* = 200 Oe applied to the in-plane direction for a Ni plate. The inset of **b** shows a close-up near the Curie temperature. **c**, In-plane *M-H* curve at 1.9, 300 and 1000 K for a SrTiO$_3$ substrate. Here, *H* was applied to the [100] direction. **d**, In-plane *M-H* curve at *M-T* curve with *H* = 2000 Oe for the oven sample holder without a sample.

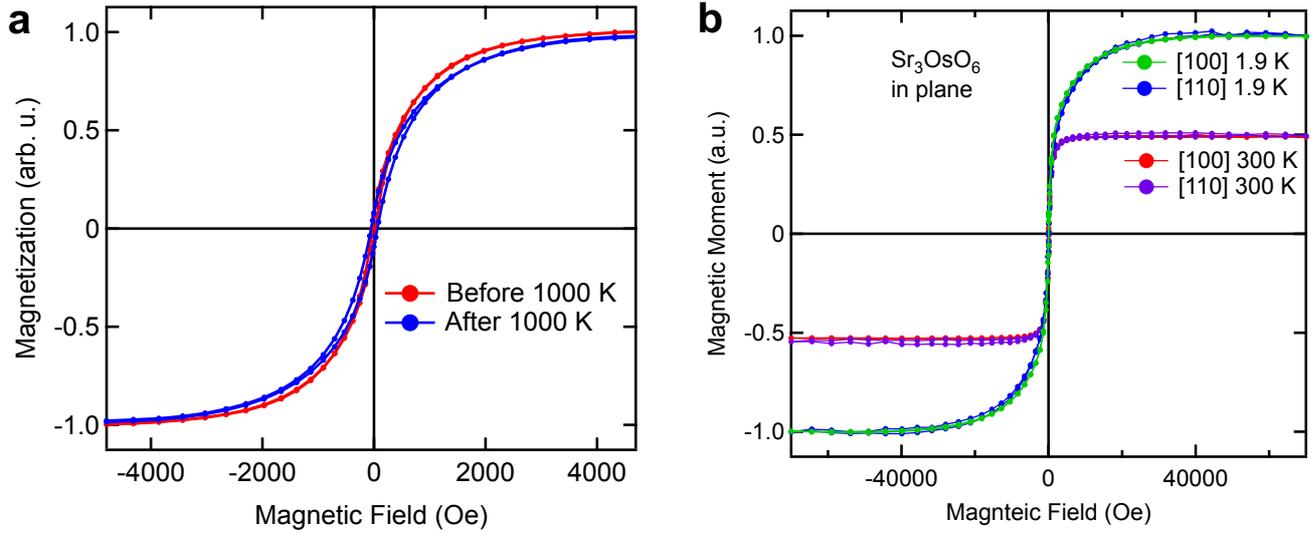

**Extended Data Figure 8 Magnetic properties of a Sr$_3$OsO$_6$ film after heating and Magnetic properties of a Sr$_3$OsO$_6$ film with *H* applied to the [100] and [110] directions. a**, In-plane *M-H* curves of a Sr$_3$OsO$_6$ film at 300 K before and after the sample was heated to 1000 K. Here, *H* was applied to the [100] direction. **b**, In-plane *M-H* curve at 1.9 and 300 K for a Sr$_3$OsO$_6$ film. Here, *H* was applied to the [100] and [110] directions.

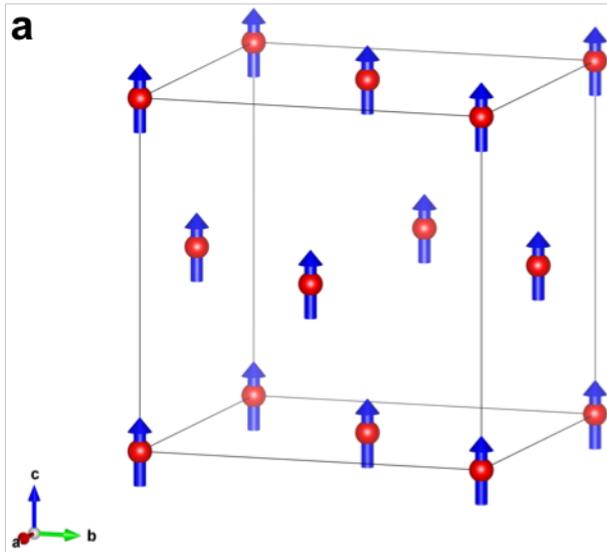 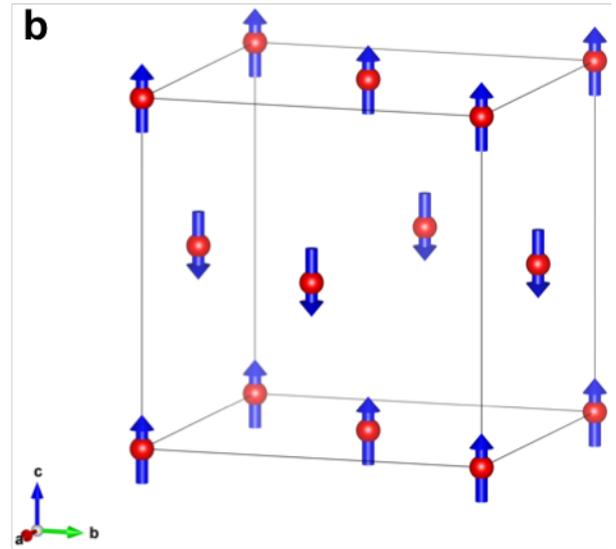

**Extended Data Figure 9 Schematic diagram of the magnetic orders. a, b,** Schematic diagram of the collinear FM order (**a**) and the AFM order (**b**). In **a** and **b**, red spheres and blue arrows indicate Os atoms and Os magnetic moments, respectively, and the Sr and O atoms are omitted for the simplicity.